# Embedding a non-diffracting defect site in helical lattice wave-field by optical phase engineering

Manish Kumar* and Joby Joseph

*Photonics Research Laboratory, Department of Physics, Indian Institute of Technology Delhi, New Delhi, India - 110016*
*Corresponding author: manishk.iitd@gmail.com*

We present a technique to optically induce a defect site in helical lattice wave-field where the combined wave-field continues to maintain its non-diffracting (ND) nature. This is done by coherently superposing a helical lattice wave-field and a Bessel beam by method of phase engineering. The results are confirmed by numerical simulations and experimentally as well by generating the ND defect beam by displaying the numerically calculated phase pattern on a phase only spatial light modulator (SLM). This technique is wavelength independent, completely scalable, and can easily be used to generate or transfer these structures in any photosensitive medium.

## 1. Introduction

Helical lattice wave-fields find many applications ranging from fabrication of chiral photonic structures [1], to the field of optical traps and micro-manipulation [2]. Helical/chiral lattices with their pitch of the order of wavelength of light have found much attention in the field of photonics due to their special ability to respond to the circular polarization state of the light and act as broadband circular polarizer [3]. Out of the methods such as, direct laser writing [4], glancing angle deposition [5], interference/holographic lithography [6] methods, for fabrication of the chiral lattices, holographic lithography based method has special advantage of being parallel in nature for large area fabrication. Recently, it was shown that these chiral photonic structures can be generated by phase engineering by employing a phase only SLM [1]. This method is essentially based on the generation of helical lattice wave-fields. Taking a parallel from the case of other photonic lattices the possibility of embedding a defect site [7] may open many new possibilities. There have been many studies on the defects in the chiral photonic lattices [8,9]. It is known for very long that lattice wave-fields are very helpful in various optical trapping applications in 2D and 3D arrays [10]. Apart from particle and biological sample trapping, helical wave-fields have also shown application in the field of cold matter trapping [11,12]. With slight modifications they may lead to micro-rotors as well [2].

Introduction of a defect site in these lattice wave-fields may bring out additional benefit of creating a channel to sort out particles as one could easily switch on or off a particular defect site in real time, by changing the displayed phase on the SLM. Thus, a selective sorting/trapping of particles/biological samples could be done. However, introduction of a ND defect site in a photonic lattice wave-field had not been done until recently when Kelberer et.al.[13] reported introduction of a defect site in the ND hexagonal lattice by coherently superimposing hexagonal lattice wave-field with zero order Bessel beam. In present work, we show how the principle of coherent addition of two ND wave-fields could be utilized to generate a ND defect site in helical lattice wave-field. We make use of an appropriate higher order Bessel beam [14] with helical lattice wave field and combine them together by means of phase engineering. We show that displaying phase part alone, via a phase only SLM, is sufficient to realize the desired wave-field. The resultant wave-filed is equivalent to removal of a single helix post from the whole lattice.

## 2. Theory and simulation

In 2011, Xavier et.al. [1] showed that a complex chiral lattice structure could be fabricated in one step by making use of 6+1 linearly polarized beams. They used a central beam symmetrically surrounded by angularly displaced six azimuthal beams, each having a designed initial phase offset given by $\Phi_i = 2\pi i/n$, where i is the index running from 1 to 6 for each of the azimuthal beams as one goes clockwise (or counterclockwise) and n = 6. The initial phase offset in each beam was introduced by the method of phase engineering using a phase only SLM. The corresponding field to generate this chiral structure is essentially a helical lattice wave-field. The phase engineering part to generate it includes firstly, finding the complex field due to interference of 6 side beams (each with specific initial offset phase) and then extracting the phase only component of this complex field numerically. This phase only component is later displayed on a phase SLM. Ideally, one should have displayed the whole complex field utilizing a combination of two SLMs [15-17] or a single phase SLM twice via a reflection from a mirror [18] which would have resulted into very efficient generation of 6 beams. However, it has been shown earlier that displaying the phase part alone on single phase SLM can also lead to efficient generation of same 6 beams. This method leads to few additional terms appearing in the Fourier transform (FT) plane which are easily removed by employing appropriate Fourier filter (FF) so that only the required terms are allowed to propagate and interfere [19,20]. While displaying the phase only component on phase SLM, we also get a significantly intense reflection at air -

SLM interface leading to a zero order term at the centre of FT plane. This zero order term may be used to contribute towards adding desired central beam along with other side beams. The amplitude of this central beam could further be increased by modifying the extracted phase function. This is done by coherent addition of a carrier plane wave, travelling perpendicular to the interfering plane, along with other 6 beams during the synthesis of phase only function. The strength of this carrier wave determines the increase in the zero order component intensity. Hence, it becomes practically possible to display the phase part alone to get 6+1 beam geometry leading to a helical lattice wave-field.

Fig. 1(a)-(b) show the numerically calculated amplitude and phase of the complex field resulting from interference of 6 side-beams. We denote this complex field by $E_S$. The inset image in Fig. 1(b) is FT of $E_S$.

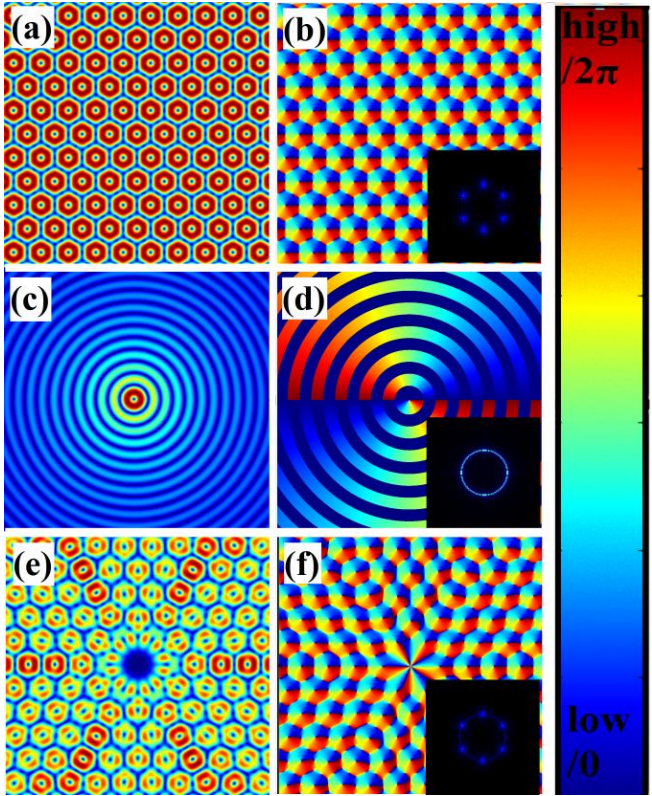

Fig. 1. (Color online) Numerical simulation results showing amplitude (left column) and phase (right column). (a)-(b) amplitude and phase of 6 beam interference field $E_S$ (with all beams having designed initial phase offset); (c)-(d) amplitude and phase of synthesized Bessel beam $E_B$; (e)-(f) amplitude and phase of resultant field $E_{Res}$. The inset images show the irradiance profile at FT plane due to each of the complex fields (amplitude and phase part together) and colorbar serves the twin purpose of representing the normalized amplitude from 0 (low) to 1 (high) and phase levels from 0 to $2\pi$. This colorbar doesn't apply to inset images.

If one displays the phase part of $E_S$, shown in Fig. 1(b), on a phase SLM, this should ideally lead to the FT pattern as shown in Fig. 2(a) which is calculated numerically by making use of fast Fourier transform (FFT) algorithm in Matlab. Therefore, an appropriate FF with transmission function as shown in Fig. 2(d) may be used to discard unwanted terms and obtain the 6 beam interference geometry. Now additional reflection at air-SLM interface gives rise to additional spot in the Fourier plane at centre (not shown in Fig.) and in that case the FF with transmission function as given in Fig. 2(f) could be utilized for obtaining a 6+1 beam interference geometry leading to ND helical lattice wave-field.

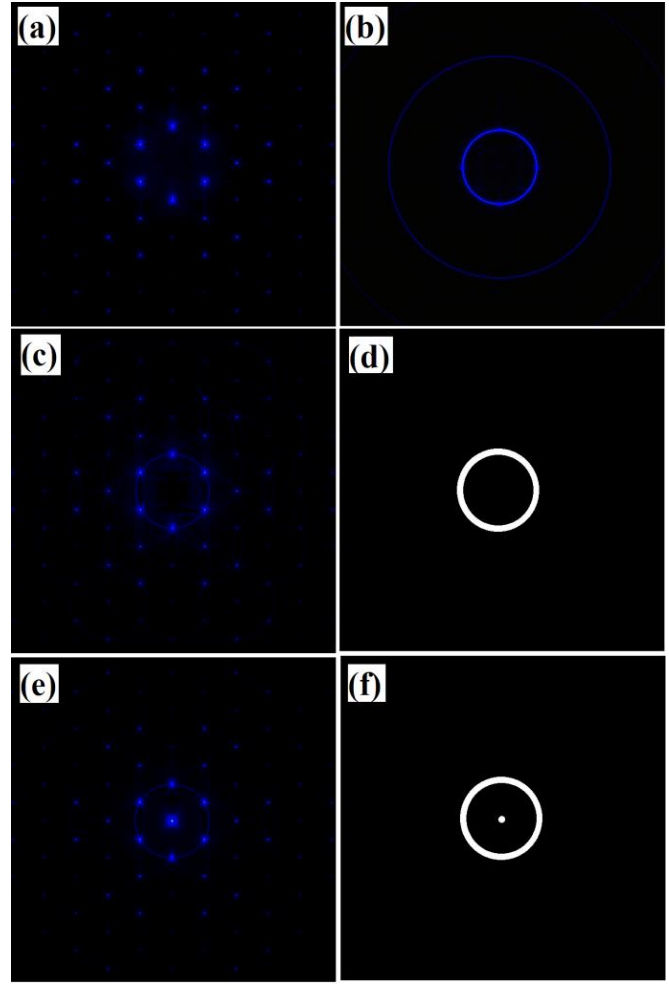

Fig. 2. (Color online) (a)-(c) shows numerically calculated irradiance profile at the Fourier plane for the phase only component of $E_S$, $E_B$ and $E_{Res}$ wave-fields respectively. (d) FF function (white = unity transmission and black = zero transmission) to discard unwanted terms from the FT of (a)-(c) to match them to corresponding FT of original complex wave-fields. (e) Numerically calculated FT of phase only component of $E_{Res}$ + carrier wave-field to give rise to zero order term. (f) FF to discard unwanted terms from (e) to get the desired wave-field.

In order to create a defect site in this ND helical lattice one need to find appropriate Bessel field which has wave-vector components pointing along a cone i.e. the FT components are distributed along a circle [13]. As the helical lattice wave-field is hollow with zero intensity in centre, a zero order Bessel beam, in spite of having its wave-vectors distributed along a cone, will not satisfy the requirement for defect creation. Here we need to employ a specific higher order Bessel beam. As the helical lattice wave-field is created by 6 azimuthal waves each having a predefined phase offset in such a way that we have a phase shift of 0 to $2\pi$ as one goes from beam index 1 to 6. So, the corresponding Bessel beam should also undergo same phase shift of 0 to $2\pi$ which is satisfied by first order Bessel beam. We also cross verify this by numerically synthesising the corresponding

Bessel field by summing the fields of very high number of side beams in umbrella geometry (n=300 in our calculation), where each of the beams have equal perpendicular component of wavevector (i.e. they make same angle of inclination with normal to the plane on which they are interfering), and have initial phase offset described by $\Phi_i = 2\pi i/n$. This synthesised field is exactly same as the first order Bessel field. The amplitude and phase part of the first order Bessel field ($E_B$) are displayed in Fig. 1(c)-(d). The FT components of this $E_B$ all lie along a cone and show up along the circumference of a circle in the FT plane as shown in the inset of Fig. 1(d). Fig. 2(b) shows numerically calculated FT due to phase part of $E_B$ alone. Now, required ND beam with defect site is realized by superimposing $E_S$ and $E_B$ coherently while maintaining a phase difference of $\pi$ between them.

$$E_{\text{Re}s} = E_S + a \cdot e^{i\pi} E_B \quad (1)$$

where factor $\exp(i\pi)$ ensures that there is destructive interference between two wave fields leading to local reduction in intensity. Here factor $a$ controls the strength of the defect and leads to zero intensity at central site as this factor tends to unity. Fig 1(e)-(f) show the amplitude and phase part of this resultant complex field ($E_{Res}$) along with its numerically calculated FT in the inset of Fig. 1(f). The FT due to phase part of $E_{Res}$ is shown in Fig. 2(c) which after accounting for the reflections at SLM interface and carrier wave addition before extraction of phase term (as explained earlier), leads to the FT intensity profile as shown in Fig. 2(e). It is very clear from the figure that here we have FT terms belonging to 6 azimuthal side beams, first order Bessel beam and the normally incident plane wave as well. There are other unwanted terms which could be removed by making use of a FF with transmission function as shown in Fig. 2(f).

As clear from the inset figures of Fig. 1, the FT components of $E_S$, $E_B$ and $E_{Res}$ wave-fields lie along a circle, of same radius, leading to the conclusion that these wave-fields have equal magnitude of perpendicular component of the wave-vectors. This makes them ND in nature and the cross-section of these wave-field remains invariant along the direction of travel of wave-field. However, when one adds a central plane wave to these wave-fields, it leads to a modulation of intensity along the direction of propagation. The actual intensity for such an arrangement of multiple interfering plane waves can be given by the general expression

$$I(r) = \sum_{i=0}^{n} |E_i|^2 + \sum_{i=0}^{n}\sum_{\substack{j=0\\j\neq i}}^{n} E_i E_j^* \cdot \exp\left[i(k_i - k_j)r + i\psi_{ij}\right] \quad (2)$$

where $E_i$, $k_{i,j}$, $r$ and $\psi_{ij}$ are the complex amplitudes, the wave vectors, the position vector, and the difference in initial offset phase of the interfering beams, respectively. Index i=0 corresponds to the central plane wave-field and i=1 to 6 describe the azimuthal wave-fields.

We use above eqn. (2) to numerically calculate the intensity profile due to interference of an on-axis plane wave with 6 azimuthal beams (each with predesigned initial offset phase). The cross-sectional view of resultant profile is shown in Fig. 3(a). To visualize the phase profile of the first order Bessel beam, an on-axis plane wave has been interfered with it, leading to a cross-sectional profile as shown in Fig. 3(b). Similarly, when we interfere an on-axis plane wave with $E_{Res}$, it leads to the intensity cross-section as shown in Fig. 3(c).

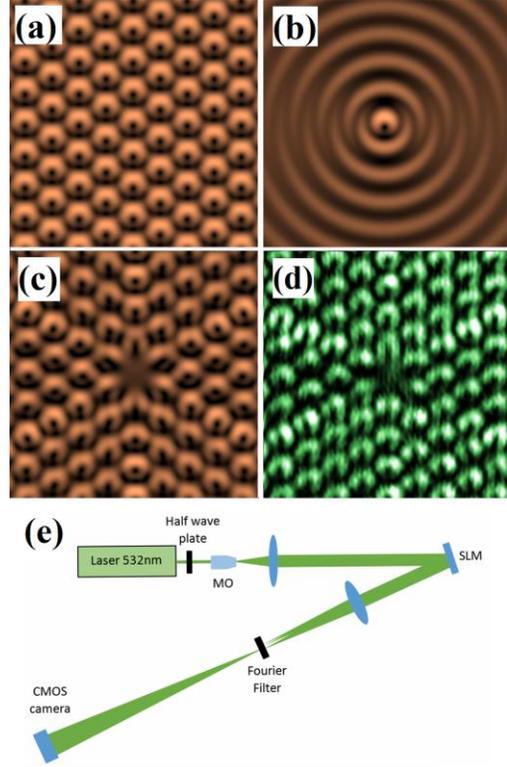

Fig. 3. (Color online) Irradiance profile at the imaging plane after proper Fourier filtering and experimental diagram. Numerically calculated results are shown in (a)-(c) for the case of 6 side + central beam interference, Bessel + central beam interference and, 6 + Bessel + central beam interference obtained according to our method respectively. Experimentally obtained result for 6+Bessel+central beam interference leading to defect embedded helical lattice wave-field is shown in (d) while schematic for experimental setup is shown in (e).

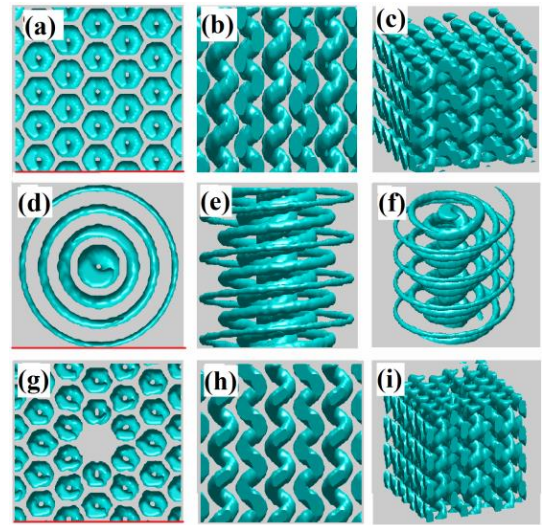

Fig. 4. (Color online) Top-view (in left column), side-view (in central column) and perspective-view (in right column) of various interference profiles (after thresholding) in space for the case of $E_S$+ on axis plane wave in (a)-(c), $E_B$+ on axis plane wave in (d)-(f), and $E_{Res}$ + on axis plane wave in (g)-(i). All the plots are obtained by applying a threshold level to actual

intensity in each layer. The red line in each of figures in left column identifies the side which is plotted in central column.

In order to appreciate the helical intensity profile we numerically calculate the intensity profiles at various depths and combine all the slices together, after thresholding the intensity at 35%, to obtain a 3D view of the corresponding intensity profiles. Fig. 4 displays the obtained 3D profiles with top, side and perspective views. One could easily conclude that wave-field continues to maintain ND behavior i.e. defect site continues throughout the depth of the wave-field.

### 3. Experimental

To verify our results experimentally we make use of a setup as described by schematic diagram shown in Fig. 3(e). We use 532 nm DPSS laser (Cobolt Samba, Sweden) a 20x microscope objective and lenses with focal length of 135mm (for collimation) and 500mm (for optical Fourier transform) and a phase only SLM. The phase only SLM is most important optical element in this experiment which is capable of introducing a change in the phase of the in-coming plane wave with least alteration to the intensity part. We use phase only SLM (Holoeye Pluto-VIS) which is a reflective LCOS micro-displays with 1920 x 1080 pixel resolution and a small 8.0 μm pixel pitch. The phase extracted from wave-field ($E_S$, $E_B$ and $E_{Res}$) were displayed on this phase only SLM and resultant irradiance profile in FT plane, same as the position of Fourier filter in Fig. 3(e), was captured using Nikon D5100 camera and it is shown here in Fig. 5. One may notice a very nice agreement with numerically calculated FT (already shown in Fig. 2).

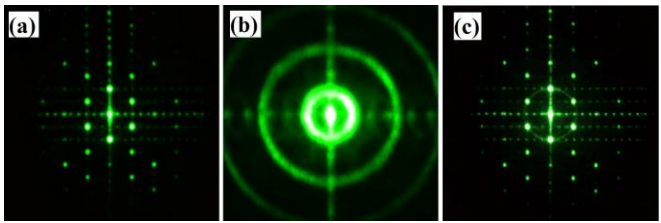

Fig. 5. (Color online) (a)-(c) Experimentally recorded irradiance profiles in the image plane due to the display of phase only components of $E_S$, $E_B$ and $E_{Res}$ on phase SLM respectively.

We notice that in FT of phase only component of $E_{Res}$, i.e. in Fig. 5(c), wave vectors belonging to $E_B$ show up as a ring. So, an appropriate FF was made by printing the FF function of Fig. 2(f) on a transparency sheet and using it in our experiment to discard the unwanted beams. Only the components belonging to $E_S$, $E_B$ and central beam were allowed by this FF and the intensity profile of resultant wave-field was captured on CMOS camera (DMK-2BUC02 from Imaging Source, Germany). The profile is presented along with corresponding numerical simulation, for easy comparison, in the Fig. 3 where we can see a nice resemblance of numerically obtained cross section profile in Fig. 3(c) with experimentally observed profile in Fig. 3(d). Still there is some noise present in the experimental result which may be accounted for because of unwanted scattering from the transparency sheet which is used as FF element. The results may be greatly improved by using a transmission SLM or fabricating a filter on optical grade glass.

The CMOS camera was moved on a translation stage manually to verify that the defect site continues throughout the depth without spreading to justify the ND characteristics. The irradiance profile captured at different depths is shown in Fig. 6 along with corresponding numerically simulated profile. It is evident that the defect site is not limited to just a single plane but is maintained throughout the depth while the lattice wave-field rotates in the usual way.

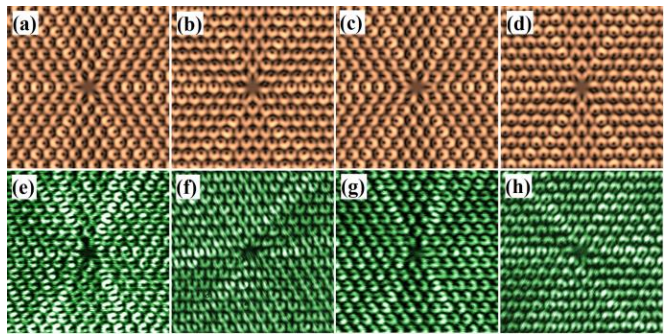

Fig. 6. (Color online) Irradiance profiles along xy-plane for different depths of z = 0, p/4, 2p/4, and 3p/4 respectively where p is the pitch of the spirals. (a)-(d) numerically simulated results and, (e)-(h) experimentally recorded images by CMOS camera in the final imaging plane. Pitch of spirals p = 1.72 cm in our case. Slow rotation of helix along with ND defect is quite evident.

In our experimental verification process we had kept the angle between side beams and surface normal to be very small (just 0.45°). This lead to helix with large pitch (1.72 centimeters) which could easily be captured by manually translating the CMOS camera on a linear translation stage. In order to reduce the pitch of helix, one could increase the angle between side beams and normal to higher values by loading the down-scaled phase function (same as zoomed out version of phase function) which is extracted from complex field formed by interference of wave-fields which make larger angle with normal. In addition, a two-lens based 4-f setup [21] (instead of single lens based 4-f setup of present work) with second lens of smaller focal length to control the demagnification factor (given by ratio of two focal lengths) could be used. Such inverted telescopic arrangement may be used to get the helix with smaller pitch and smaller periodicity for actual micro-fabrication or trapping application. As already pointed out, the use of a transmission SLM instead of printed transparency sheet as FF would do a lot of improvement in the quality of result for a real application.

### 4. Conclusion

We have presented a method to incorporate a local ND defect site in helical lattice wave-field by coherent superposition of helical lattice wave-field with first order Bessel beam where the wave-field continues to maintain its ND characteristics. This is done by making use of a single phase only SLM. Such a beam could be useful for micro fabrication of defects in chiral photonic crystals or in many optical trapping/micromanipulations related application.

### 5. Acknowledgement

Manish Kumar wishes to acknowledge the financial support from Council of Scientific and Industrial Research (CSIR), India.